\documentclass[fp,twocolumn,amsmath]{jpsj3}

\usepackage{amsmath}
\usepackage{txfonts}
\usepackage{bm}

\usepackage{color}

\newcommand{\bra}  {\langle}
\newcommand{\ket}  {\rangle}

\title{Self-doping effect arising from electron correlations in multi-layer cuprates}

\author{Kazutaka Nishiguchi\thanks{nishiguchi@mp.es.osaka-u.ac.jp}, Shingo Teranishi, Koichi Kusakabe} 
\inst{Graduate School of Engineering Science, Osaka University, \\ 1-3 Machikaneyama-cho, Toyonaka, Osaka, 560-8531, Japan}

\abst{
A self-doping effect between outer and inner CuO$_2$ planes (OPs and IPs) in multi-layer cuprate superconductors is studied. 
When one considers a three-layer tight-binding model of the Hg-based three-layer cuprate derived from the first principle calculations, 
the electron concentration gets to be large in the OP compared to IP. 
This is inconsistent with the experimental fact that more hole carriers tend to be introduced into the OP than IP. 
We investigate a three-layer Hubbard model with the two-particle self-consistent approach for multi-layer systems 
to incorporate electron correlations. 
We observe that the double occupancy (antiferromagnetic instability) in the IP decreases (increases) more than the OP, 
and also reveal that more electrons tend to be introduced into the IP than OP to obtain the energy gain from the on-site Hubbard interaction. 
These results are consistent with the experimental facts, 
and this electron distribution between the OP and IP can be interpreted as a self-doping effect arising from strong electron correlations. 
}

\begin{document}
\maketitle

\section{Introduction}

Since the discovery of cuprate superconductors, 
high-$T_{\text{c}}$ (critical temperature) superconductivity has attracted much interests 
both experimentally and theoretically in condensed matter physics.~\cite{Scalapino12,Aoki12} 
Although many parent compounds of cuprates are Mott insulators, 
the high-$T_{\text{c}}$ superconducting (SC) phase 
and the other novel quantum phases such as antiferromagnetic (AF), pseudogap, and non-Fermi liquid phase 
appear once mobile carriers are introduced into the CuO$_2$ plane, 
where their quantum phenomena arise from competition of the itinerancy and localization of electrons due to electron correlations. 
Among various families of superconductors, 
cuprates still stand out in having the highest-$T_{\text{c}}$ superconductor under normal pressures to date, 
and the highest-$T_{\text{c}}$ occurs in multi-layer cuprate superconductors that possess $n$ CuO$_2$ planes in a unit cell, 
typically Hg-based cuprates HgBa$_2$Ca$_{n-1}$Cu$_n$O$_{2n+2+\delta}$ [Hg-12$(n-1)n$] ($\delta$ is the chemical doping rate coming from oxygens). 
In the multi-layer cuprate superconductors the $T_{\text{c}}$ depends on the number $n$ of the CuO$_2$ planes:~\cite{Leggett06TB,Mukuda12} 
the $T_{\text{c}}$ increases for $1 \leq n \leq 3$ and decreases slightly and saturates for $n \geq 3$, 
and the Hg-based three-layer cuprate Hg-1223 is the highest $T_{\text{c}}$ ($\sim 135$ K) superconductor among cuprates.~\cite{Schilling93,Yamamoto15} 

Many theoretical attempts have proposed 
several mechanisms of high-$T_{\text{c}}$ superconductivity in the multi-layer cuprates: 
they have suggested 
an inter-layer Josephson coupling arising from a second-order process of the inter-layer single-electron hopping~\cite{Anderson97TB,Chakravarty93}, 
an inter-layer Josephson pair tunneling in a macroscopic Ginzburg-Landau free energy scheme,~\cite{Chakravarty04} 
a Coulomb energy saving in the $c$-axis layering structure,~\cite{Leggett98,Leggett99} 
% while Okamoto and Maier studied an effect of the interlayer one-electron hopping for double-layer Hubbard and $t$-$J$ models. 
% Chen et al. have also examined an effect of the interlayer tunneling in terms of a free energy derived from the $t$-$J$ model 
% in a case where the phenomenological interlayer coupling is chosen to realize the in-phase gap function between the two layers. 
and an inter-layer pair hopping processes arising from the higher-order processes of the Coulomb interaction.~\cite{Kusakabe09,Kusakabe12,KN13} 
In addition to this, 
several nuclear magnetic resonance (NMR) experiments have shown 
the elaborate experimental results for the outer and inner CuO$_2$ planes (OPs and IPs) in multi-layer cuprate superconductors, 
where such asymmetric CuO$_2$ planes OPs and IPs appear for the number of CuO$_2$ planes $n \geq 3$. 
The NMR experiments~\cite{Mukuda12,Mukuda08,Mukuda06,Kotegawa01} have observed different properties between the OPs and IPs: 
the AF moments in the IPs is much larger than the OPs, 
and even around the optimal-doped regions the AF phase coexists with the SC phase in the IPs. 
On the other hand, 
the SC gaps with another critical temperature $T^{\prime}_{\text{c}}$ lower than bulk $T_{\text{c}}$ determined from the OPs develop 
in the IPs due to the proximity effect, 
and the superconductivity in the OPs precedes with increasing the carrier concentration until the over-doped regions. 
Furthermore, the carrier concentration is different between the OPs and IPs, i.e., 
more hole carriers tend to be introduced into the OPs than IPs. 
These different behaviors between the OPs and IPs cannot be understood microscopically, 
and theoretical interpretations starting from the microscopic models are still awaited. 

To clarify these phenomena in the multi-layer cuprates, 
in this paper we concentrate on the Hg-based three-layer cuprate Hg-1223 
as a typical example for the multi-layer cuprate superconductors, 
which includes asymmetric CuO$_2$ planes: two OPs and one single IP. 
To understand differences of electronic properties among materials, 
it is useful to start from the first principle calculations based on the density functional theory (DFT). 
Although the Mott insulating phase cannot be described by the DFT band calculations, 
they give us a good basis set for considering strongly correlated systems. 
As shown later, 
we can obtain a three-layer tight-binding model for Hg-1223 
from the DFT band calculations and maximally localized wannier functions. 
However, 
this effective model shows that the electron concentration gets to be large in the OP compared to IP. 
This is an opposite result to the experimental fact that more hole carriers tend to be introduced into the OP than IP. 

Motivated above, 
to incorporate electron correlations, 
a three-layer Hubbard model is investigated with the two-particle self-consistent approach for multi-layer systems. 
We observe that the double occupancy (AF instability) in the IP decrease (increases) more than OP. 
We also reveal that more electrons tend to be introduced into the IP than OP with increasing the on-site Hubbard interaction 
in order to obtain the energy gain from the on-site Hubbard interaction. 
These results are consistent with the experimental fact suggested above, 
and the electron distribution between the OP and IP can be interpreted as a self-doping effect arising from electron correlations.

\section{Formalism} 
In this study a three-layer Hubbard model is investigated with multi-layer TPSC approach. 
The three-layer Hubbard is given as an effective model for Hg-based three-layer cuprate Hg-1223, 
and then their parameters are also obtained from maximally localized wannier functions derived from band structures of Hg-1223. 
To investigate such a multi-layer Hubbard model, 
we consider the TPSC approach for multi-layer systems, 
and evaluate the interacting Green's function from the self-energy with the TPSC approach.

\subsection{Three-layer Hubbard model} 
Let us consider an effective model for Hg-1223 derived from the first principle calculations, 
i.e., a three-layer Hubbard model: 
\begin{equation} 
H = H_{0} +H_{\text{int}} . 
\end{equation} 
The total Hamiltonian $H$ is composed of the one-body part $H_{0}$ and two-body one $H_{\text{int}}$. 
$H_{0}$ describes a three-layer tight-binding model, 
\begin{equation} 
H_{0} = -\sum_{ij} \sum_{\sigma} \sum_{ab} t^{ab}_{ij} c^{a\dagger}_{i\sigma} c^{b}_{j\sigma} -\mu \sum_{i\sigma a} n^{a}_{i\sigma} , 
\end{equation} 
where $c^{a\dagger}_{i\sigma}$ ($c^{a}_{i\sigma}$) is the field operator 
which creates (annihilates) an electron at site $i$ with spin $\sigma \, (=\uparrow, \downarrow)$ in layer $a \, (=1,2,3)$, 
and $n^{a}_{i\sigma} = c^{a\dagger}_{i\sigma} c^{a}_{i\sigma}$.  
Here $t^{ab}_{ij}$ represents the (spin-independent) transfer integral of single-electron hopping to $(i,a)$ from $(j,b)$, 
and $\mu$ denotes the chemical potential. 
We also specify layer $a=1,3$ as two OPs and $a=2$ as one single IP. 
In this paper, 
the intra-layer single-electron hopping is taken into account until the third-neighbor hopping, 
and the inter-layer single-electron hopping works between the OP and IP only. 
Furthermore, $H_{\text{int}}$ represents the Coulomb interaction, 
\begin{equation} 
H_{\text{int}} = U\sum_{ia} n^{a}_{i\uparrow} n^{a}_{i\downarrow} , 
\end{equation} 
which means the on-site Hubbard interaction $U$. 

The one-body part $H_{0}$ can be rewritten by using the Fourier transformation for the field operators 
$c^{a}_{\bm{k}\sigma} = (1/\sqrt{N}) \sum_i \mathrm{e}^{-i\bm{k} \cdot \bm{R}_i} c^{a}_{i\sigma}$, 
where $N$ is the number of sites and $\bm{R}_i$ represents the position of site $i$ on the square lattice. 
Then $H_{0}$ can be given as a $3 \times 3$ matrix form 
% \begin{equation} 
% H_{0} 
% = \sum_{\bm{k} \sigma} 
%   \begin{pmatrix}  c^{1\dagger}_{\bm{k}\sigma} \ c^{2\dagger}_{\bm{k}\sigma} \ c^{3\dagger}_{\bm{k}\sigma} \end{pmatrix} 
%   \begin{pmatrix}  \xi_{\bm{k}} &  t_{\bm{k}}   &  0                  \\ 
%                    t_{\bm{k}}   &  \xi_{\bm{k}} &  t_{\bm{k}}  \\ 
%                    0            &  t_{\bm{k}}   &  \xi_{\bm{k}}       \end{pmatrix} 
%   \begin{pmatrix}  c^{1}_{\bm{k}\sigma}  \\  c^{2}_{\bm{k}\sigma}  \\  c^{3}_{\bm{k}\sigma} \end{pmatrix} 
% \end{equation} 
\begin{equation} 
\begin{split} 
H_{0} 
&=  \sum_{\bm{k}\sigma} \vec{c}^{\, \dagger}_{\bm{k}\sigma} \hat{\xi}_{\bm{k}} \vec{c}_{\bm{k}\sigma}  \\ 
&=  \sum_{\bm{k}\sigma}
    \left( \begin{array}{ccc} c^{1\dagger}_{\bm{k} \sigma} & c^{2\dagger}_{\bm{k} \sigma} & c^{3\dagger}_{\bm{k} \sigma} \end{array} \right)  \\ 
&\qquad \times 
    \left( \begin{array}{ccc}  \epsilon_{\bm{k}}-\mu  &  t_{\bm{k}}    &  0             \\
                               t_{\bm{k}}    &  \epsilon_{\bm{k}}-\mu  &  t_{\bm{k}}    \\
                               0             &  t_{\bm{k}}    &  \epsilon_{\bm{k}}-\mu  \end{array} \right) 
    \left( \def\arraystretch{1.1} \begin{array}{c}  c^{1}_{\bm{k} \sigma} \\  c^{2}_{\bm{k} \sigma} \\  c^{3}_{\bm{k} \sigma} \end{array} \right) , 
    \label{eq:H0} 
\end{split} 
\end{equation} 
% \begin{equation} 
% \begin{split} 
% H_{0} 
% &=  \sum_{\bm{k}\sigma} \vec{c}^{\, \dagger}_{\bm{k}\sigma} \hat{\xi}_{\bm{k}} \vec{c}_{\bm{k}\sigma}  \\ 
% &=  \sum_{\bm{k}\sigma}
%     \begin{pmatrix}  c^{1\dagger}_{\bm{k} \sigma} & c^{2\dagger}_{\bm{k} \sigma} & c^{3\dagger}_{\bm{k} \sigma} \end{pmatrix} 
%     \begin{pmatrix}  \epsilon_{\bm{k}}-\mu  &  t_{\bm{k}}             &  0             \\
%                      t_{\bm{k}}             &  \epsilon_{\bm{k}}-\mu  &  t_{\bm{k}}    \\
%                      0                      &  t_{\bm{k}}             &  \epsilon_{\bm{k}}-\mu  \end{pmatrix} 
%     \left( \def\arraystretch{1.1} \begin{array}{c}  c^{1}_{\bm{k} \sigma} \\  c^{2}_{\bm{k} \sigma} \\  c^{3}_{\bm{k} \sigma} \end{array} \right) , 
% \end{split} 
% \end{equation} 
where the field operators in a vector form represent 
$\vec{c}^{\, \dagger}_{\bm{k}\sigma} = (c^{1\dagger}_{\bm{k} \sigma} \, c^{2\dagger}_{\bm{k} \sigma} \, c^{3\dagger}_{\bm{k} \sigma})$ and 
$\vec{c}_{\bm{k}\sigma} = (c^{1}_{\bm{k} \sigma} \, c^{2}_{\bm{k} \sigma} \, c^{3}_{\bm{k} \sigma})^{\mathrm{T}}$, 
and $\xi$ indicates the energy dispersion matrix defined by the second line above.  
Here $\epsilon_{\bm{k}} -\mu$ is the intra-layer energy dispersion measured from the chemical potential $\mu$, 
\begin{equation} 
\begin{split} 
\epsilon_{\bm{k}} 
&= -2t                 \left( \cos  k_{x} +\cos  k_{y} \right)  \\ 
&\quad 
   +4t^{\prime}               \cos k_{x} \cos k_{y} 
   -2t^{\prime \prime} \left( \cos 2k_{x} +\cos 2k_{y} \right) , 
\end{split} 
\end{equation} 
and $t_{\bm{k}}$ is the inter-layer single-electron hopping, 
\begin{equation} 
t_{\bm{k}} = -t_{\perp} \left( \cos  k_{x} -\cos  k_{y} \right)^{2} . 
\end{equation} 

Now $t$, $t^{\prime}$, and $t^{\prime \prime}$ represent 
the intra-layer transfer integral for the nearest, second, third neighbor hopping, respectively. 
The inter-layer single-electron hopping can be described by one single parameter $t_{\perp}$.~\cite{Andersen94,Andersen95,KN13} 
The downfolded parameters for Hg-based cuprate Hg-1223 derived from the first principle calculations~\cite{KN_thesis} are 
$( t, t^{\prime}, t^{\prime \prime}, t_{\perp} ) = ( 0.45, 0.10, 0.08, 0.05 )$ eV. 
Also, the other Hg-based multi-layer cuprates Hg-$12(n-1)n$ also have very similar parameters to them.~\cite{KN_thesis} 
For simplicity we have omitted the difference of the site potential between the OP and IP. 
According to the first principle calculations, 
the site potential in the IP $\varepsilon_{\text{IP}}$ is larger than that in the OP $\varepsilon_{\text{OP}}$, 
and its difference $\Delta \varepsilon = \varepsilon_{\text{IP}} -\varepsilon_{\text{OP}} \sim 0.1$ eV.

% We can diagonalize the one-body part $H_{0}$ immediately 
% and obtain the energy eigenvalues 
% $\hat{E}_{\bm{k}} = \mathrm{diag} \, ( E^{1}_{\bm{k}}, E^{2}_{\bm{k}}, E^{3}_{\bm{k}} ) 
%                   = \mathrm{diag} \, \left( \xi_{\bm{k}}-\sqrt{2}t_{\bm{k}}, \, \xi_{\bm{k}}, \, \xi_{\bm{k}}+\sqrt{2}t_{\bm{k}} \right)$, 
% where we have here specified $E^{1}_{\bm{k}} \geq E^{2}_{\bm{k}} \geq E^{3}_{\bm{k}}$ due to $t_{\bm{k}} \leq 0$. 
We can diagonalize the one-body part $H_{0}$ immediately: 
the energy eigenvalues $E^{m}_{\bm{k}}$ ($m=1,2,3$) can be specified as 
% $( E^{1}_{\bm{k}}, \, E^{2}_{\bm{k}}, \, E^{3}_{\bm{k}} ) 
%  =  \left( \xi_{\bm{k}}-\sqrt{2}t_{\bm{k}}, \, \xi_{\bm{k}}, \, \xi_{\bm{k}}+\sqrt{2}t_{\bm{k}} \right)$, 
$E^{1,3}_{\bm{k}} = \xi_{\bm{k}} \mp \sqrt{2}t_{\bm{k}}$ and $E^{2}_{\bm{k}} = \xi_{\bm{k}}$, 
where we have here specified $E^{1}_{\bm{k}} \geq E^{2}_{\bm{k}} \geq E^{3}_{\bm{k}}$ due to $t_{\bm{k}} \leq 0$. 
The corresponding field operators $a^{m}_{\bm{k}\sigma}$ ($m=1,2,3$) can be also given as 
% $( a^{1}_{\bm{k}\sigma}, \, a^{2}_{\bm{k}\sigma}, \, a^{3}_{\bm{k}\sigma} ) 
%  =  \left( ( c^{1}_{\bm{k}\sigma} -\sqrt{2}c^{2}_{\bm{k}\sigma} +c^{3}_{\bm{k}\sigma} )/2, \, 
%            ( c^{1}_{\bm{k}\sigma}                               -c^{3}_{\bm{k}\sigma} )/\sqrt{2}, \, 
%            ( c^{1}_{\bm{k}\sigma} +\sqrt{2}c^{2}_{\bm{k}\sigma} +c^{3}_{\bm{k}\sigma} )/2 \right)$. 
$a^{1,3}_{\bm{k}\sigma} 
 = ( c^{1}_{\bm{k}\sigma} \mp \sqrt{2}c^{2}_{\bm{k}\sigma} +c^{3}_{\bm{k}\sigma} )/2$ 
and $a^{2}_{\bm{k}\sigma} 
 = ( c^{1}_{\bm{k}\sigma}                               -c^{3}_{\bm{k}\sigma} )/\sqrt{2}$. 
In this study the carrier doping can be adjusted by the chemical potential $\mu$.

\subsection{TPSC approach for multi-layer systems} 
We here show the procedure of the TPSC approach for multi-layer systems, 
i.e., multi-layer TPSC approach. 
The TPSC approach~\cite{Vilk94,Vilk97,Arita00,Miyahara13} is one of the weak- and intermediate-coupling theories, 
where the conservation law of spin and charge, Mermin-Wagner theorem, 
Pauli principle, $q$-sum rule for spin and charge susceptibility, and $f$-sum rule are satisfied. 
In the TPSC approach, 
the spin and charge susceptibility are determined self-consistently together with the double occupancy by assuming the TPSC ansatz, 
and then the self-energy and Green's function can be evaluated straightforwardly from them. 
From now on, we develop the TPSC approach to apply to multi-layer systems including the three-layer Hubbard model. 

To understand the one-particle electronic properties, 
we evaluate the Green's function in the multi-layer systems, defined as 
\begin{equation} 
G^{ab}(k) = -\int^{\beta}_{0} d\tau \, \mathrm{e}^{i\omega_{n}\tau} \bra T_{\tau} c^{a}_{\bm{k}\sigma}(\tau) c^{b\dagger}_{\bm{k}\sigma}(0) \ket , 
\end{equation} 
where $k=(\bm{k}, i\omega_{n})$ and the Matsubara frequency for Fermions $\omega_{n} = (2n+1)\pi /\beta$ ($n \in \bm{Z}$). 
To obtain this with the multi-layer TPSC approach, 
let us first consider the spin and charge (orbital) susceptibility in the multi-layer systems. 
Using the spin operators in the momentum space 
$S^{z \, a}_{\bm{q}}= (1/2) \sum_{\bm{k}} [ c^{a\dagger}_{\bm{k}\uparrow  } c^{a}_{\bm{k}+\bm{q}\uparrow  } 
                                           -c^{a\dagger}_{\bm{k}\downarrow} c^{a}_{\bm{k}+\bm{q}\downarrow} ]$, 
$S^{- \, a}_{\bm{q}}= \sum_{\bm{k}} c^{a\dagger}_{\bm{k}\downarrow} c^{a}_{\bm{k}+\bm{q}\uparrow}$, 
and 
$S^{+ \, a}_{\bm{q}}= \sum_{\bm{k}} c^{a\dagger}_{\bm{k}\uparrow} c^{a}_{\bm{k}+\bm{q}\downarrow}$ at layer $a$, 
and the charge susceptibility in the momentum space 
$n^{a}_{\bm{q}}= \sum_{\bm{k}} [ c^{a\dagger}_{\bm{k}\uparrow  } c^{a}_{\bm{k}+\bm{q}\uparrow  } 
                                +c^{a\dagger}_{\bm{k}\downarrow} c^{a}_{\bm{k}+\bm{q}\downarrow} ]$ at layer $a$, 
we define the longitudinal and transverse spin susceptibility as 
\begin{equation} 
\begin{split} 
\chi^{ab}_{\text{S}zz}(q) 
&=  \frac{1}{N} \int^{\beta}_{0} d\tau \, \mathrm{e}^{i\epsilon_{m} \tau} \bra T_{\tau} S^{z \, a}_{\bm{q}} (\tau) S^{z \, b}_{-\bm{q}} (0) \ket ,  \\ 
\chi^{ab}_{\text{S}\pm}(q) 
&=  \frac{1}{N} \int^{\beta}_{0} d\tau \, \mathrm{e}^{i\epsilon_{m} \tau} \bra T_{\tau} S^{- \, a}_{\bm{q}} (\tau) S^{+ \, b}_{-\bm{q}} (0) \ket , 
\end{split} 
\end{equation} 
and the charge (orbital) susceptibility as 
\begin{equation} 
\begin{split} 
\chi^{ab}_{\text{C}}(q) 
&=  \frac{1}{N} \int^{\beta}_{0} d\tau \, \mathrm{e}^{i\epsilon_{m} \tau}  
    \frac{1}{2} [ \bra T_{\tau} n^{a}_{\bm{q}} (\tau) n^{b}_{-\bm{q}} (0) \ket -\bra n^{a}_{\bm{q}} \ket \bra n^{b}_{-\bm{q}} \ket ] ,
\end{split} 
\end{equation} 
where $q=(\bm{q}, i\epsilon_{m})$, the Matsubara frequency for Bosons $\epsilon_{m} = 2m\pi /\beta$ ($m \in \bm{Z}$), 
and the inverse temperature $\beta = 1/T$ ($k_{\text{B}}=1$). 
Here $T_{\tau}$ indicates the imaginary-time ordered product 
and $\bra \cdots \ket$ represents the quantum statistical average. 
In the presence of spin SU(2) symmetry, 
the longitudinal and transverse spin susceptibility satisfy $\hat{\chi}_{\text{S}} \equiv 2\hat{\chi}_{\text{S}zz} = \hat{\chi}_{\text{S}\pm}$. 

In the multi-layer TPSC approach 
the spin and charge (orbital) susceptibility are assumed to take the following forms:  
\begin{equation} 
\begin{split} 
\hat{\chi}_{\text{S}}(q) = \frac{\hat{\chi}_{0}(q)}{1 -\hat{\chi}_{0}(q) \hat{U}_{\text{S}}} , 
\quad 
\hat{\chi}_{\text{C}}(q) = \frac{\hat{\chi}_{0}(q)}{1 +\hat{\chi}_{0}(q) \hat{U}_{\text{C}}} . 
\end{split} 
\end{equation} 
Note that the matrix products are defined here as 
$\frac{\hat{\chi}_{0}}{1 -\hat{\chi}_{0} \hat{U}_{\text{S}(\text{C})}} = [1 -\hat{\chi}_{0} \hat{U}_{\text{S}(\text{C})}]^{-1} \hat{\chi}_{0}$. 
Here the polarization function $\chi_{0}$ is given as 
\begin{equation} 
\chi^{ab}_{0}(q) = -\frac{1}{N\beta} \sum_{k} G^{ab}_{0}(q+k) G^{ba}_{0}(k) , 
\end{equation} 
obtained from the non-interacting Green's function $\hat{G}_{0}(k) = ( i\omega_{n} -\hat{\xi}_{\bm{k}} )^{-1}$. 
The spin and charge channel interaction, 
\begin{equation} 
\begin{split} 
\hat{U}_{\text{S}} 
&=  \left( \begin{array}{ccc}  U^{\text{OP}}_{\text{S}}  &  0                         &  0  \\
                               0                         &  U^{\text{IP}}_{\text{S}}  &  0  \\
                               0                         &  0                         &  U^{\text{OP}}_{\text{S}}  \end{array} \right) , \quad  \\ 
\hat{U}_{\text{C}} 
&=  \left( \begin{array}{ccc}  U^{\text{OP}}_{\text{C}}  &  0                         &  0  \\
                               0                         &  U^{\text{IP}}_{\text{C}}  &  0  \\
                               0                         &  0                         &  U^{\text{OP}}_{\text{C}}  \end{array} \right) , 
\end{split} 
\end{equation} 
are determined self-consistently together with the double occupancy 
$\bra n^{a}_{\uparrow} n^{a}_{\downarrow} \ket \equiv \bra n^{a}_{i\uparrow} n^{a}_{i\downarrow} \ket$
by the $q$-sum rule for the spin and charge susceptibility and TPSC ansatz in multi-layer systems: 
\begin{equation} 
\begin{split} 
\frac{1}{N\beta} \sum_{q} 2\chi^{aa}_{\text{S}}(q) &= n^{a} -2\bra n^{a}_{\uparrow} n^{a}_{\downarrow} \ket ,  \\ 
\frac{1}{N\beta} \sum_{q} 2\chi^{aa}_{\text{C}}(q) &= n^{a} +2\bra n^{a}_{\uparrow} n^{a}_{\downarrow} \ket -(n^{a})^{2} , 
\end{split} 
\end{equation} 
and 
\begin{equation} 
U^{aa}_{\text{S}} = U \frac{\bra n^{a}_{\uparrow} n^{a}_{\downarrow} \ket}{\bra n^{a}_{\uparrow} \ket \bra n^{a}_{\downarrow} \ket} .  \label{eq:ansatz} 
\end{equation} 
Here the layer filling $n^{a} = \bra n^{a}_{\uparrow} +n^{a}_{\downarrow} \ket$ at layer $a$ is 
the $\sigma$-summation of the electron concentration 
$\bra n^{a}_{\sigma} \ket = \bra n^{a}_{i\sigma} \ket$ at layer $a$ with spin $\sigma$. 
Now we assume paramagnetic states so that $\bra n^{a}_{\uparrow} \ket = \bra n^{a}_{\downarrow} \ket$. 

After determining the spin and charge (orbital) susceptibility, 
we can straightforwardly obtain the following expression for the self-energy from them, as  
\begin{equation} 
\begin{split} 
\Sigma^{ab}(k) 
&=  \frac{1}{N\beta} \sum_{k^{\prime}} 
    \bigg[ \hat{U} +\frac{3}{4} \hat{U} \hat{\chi}_{\text{S}}(k-k^{\prime}) \hat{U}_{\text{S}}  \\ 
&\qquad \qquad \qquad 
                   +\frac{1}{4} \hat{U} \hat{\chi}_{\text{C}}(k-k^{\prime}) \hat{U}_{\text{C}} \bigg]^{ab} G^{ab}_{0}(k^{\prime}) ,  \label{eq:self-energy} 
\end{split} 
\end{equation} 
where $\hat{U} = \mathrm{diag} \, (U, \, U, \, U)$ denotes the ``bare" on-site Hubbard interaction. 
Thus the interacting Green's function in the multi-layer TPSC approach can be given straightforwardly as 
\begin{equation} 
\hat{G}(k) = \Big[ \hat{G}^{-1}_{0}(k) -\hat{\Sigma}(k) \Big]^{-1} . 
\end{equation} 

To evaluate the layer filling $n^{a}$ at layer $a$, 
and also to determine the chemical potential from the total filling, 
it is useful to obtain $n^{a}$ from the Green's function as follows: 
\begin{equation} 
n^{a} = \frac{1}{N} \sum_{\bm{k}\sigma} \left( \frac{2}{\beta} \sum_{\omega_{n} >0} \mathrm{Re} \, G^{aa}(\bm{k},i\omega_{n}) +\frac{1}{2} \right) . 
\end{equation} 
This is a general expression satisfied for both interacting and non-interacting case.

\section{Numerical Results} 
The numerical results are shown as follows: 
the non-interacting results where the three-layer tight-binding model (only the one-body part) is taken into account, 
and the TPSC results where the three-layer Hubbard model (with the on-site Hubbard interaction) is investigated with the multi-layer TPSC approach. 
In our numerical results 
we commonly set the number of the discrete mesh points for the wave number and Matsubara frequency to be  
$(k_{x}, k_{y}, \omega_{n} (\epsilon_{m})) = (64, 64, 4096)$, and the temperature to be $T=0.01$ eV ($\sim 100$ K). 

\subsection{Non-interacting results} 
\begin{figure}[t]
\begin{center}
\includegraphics[width=8.0cm,clip]{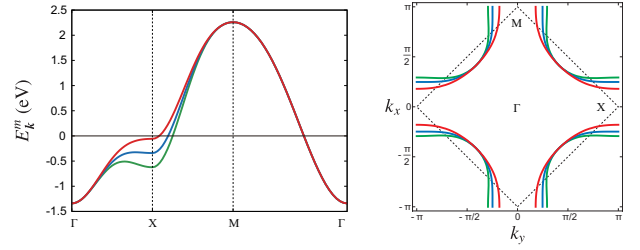} 
\caption{ 
(Left) Energy band structure measured from the chemical potential at the half-filling $n_{\text{av}}=1$ 
and (Right) its Fermi surfaces of the three-layer tight-binding model. 
The Fermi surface for perfect nesting is also depicted here.
} 
\label{fig:BAND_FS}
\end{center} 
\end{figure}

Let us start with the non-interacting results obtained from the three-layer tight-binding model. 
We here treat the average filling $n_{\text{av}}= (1/3) \sum^{3}_{a=1} n^{a}$ as a parameter, 
which can be specified by adjusting the chemical potential $\mu$. 
The energy band structure measured from $\mu$ at the half-filling $n_{\text{av}}=1$ 
and its Fermi surfaces of the three-layer tight-binding model 
are shown in Fig. \ref{fig:BAND_FS}. 
The energy bands along the $M$-$\Gamma$ line are triply degenerated due to $t_{\bm{k}}=0$, 
while those around the $X$ point are split due to $t_{\bm{k}} \neq 0$. 

\begin{figure}[t] 
\begin{center} 
\includegraphics[width=8.0cm,clip]{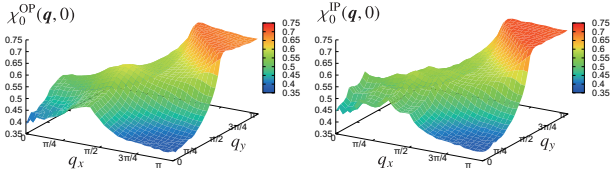} 
\caption{ 
Polarization function in (Left) OP and (Right) IP ($0 \leq q_{x,y} \leq \pi$). 
The parameters are set to be $n_{\text{av}}=0.85$ and $\epsilon_{m}=0$. 
} 
\label{fig:X0} 
\end{center} 
\end{figure}

In Fig. \ref{fig:X0} we next show the polarization function in the OP and IP ($0 \leq q_{x,y} \leq \pi$), 
which are defined as $\chi^{\text{OP}}_{0} \equiv \chi^{11}_{0} = \chi^{33}_{0}$ 
and $\chi^{\text{IP}}_{0} \equiv \chi^{22}_{0}$, respectively. 
Here we have set the following parameters: 
the average filling $n_{\text{av}}=0.85$ and zero (Matsubara) frequency $\epsilon_{m}=0$. 
One can observe that both $\chi^{\text{OP}}_{0}$ and $\chi^{\text{IP}}_{0}$ have peaks near the nesting vector $\bm{Q}=( \pm \pi, \pm \pi  )$, 
and that the peak for $\chi^{\text{IP}}_{0}$ is larger than $\chi^{\text{OP}}_{0}$. 
Therefore the AF fluctuation is large both in the OP and IP, 
and the IP have strong AF instability compared to the OP. 
This is qualitatively consistent with the experimental fact that the IP have large AF moment compared to the OP. 

\begin{figure}[t] 
\begin{center} 
\includegraphics[width=8.0cm,clip]{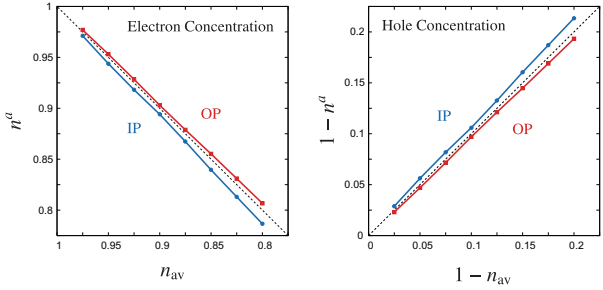} 
\caption{ 
(Left) Electron concentration 
and (Right) hole concentration with varying the average filling $n_{\text{av}}$.
} 
\label{fig:lf0}
\end{center} 
\end{figure}

On the other hand, 
we can compare the electron (hole) concentration $n^{a}$ ($1-n^{a}$) in the OP and IP, 
which is shown in Fig. \ref{fig:lf0} with varying the average filling $n_{\text{av}}$ ($1-n_{\text{av}}$). 
One can see that in each case the electron concentration is larger in the OP than IP, 
so that more hole carriers tend to be introduced into the IP than OP. 
This result may be understood naturally as follows. 
If the hole carriers were introduced into the energy bands, 
they would start to be introduced into the highest energy band ($m=1$) having the energy band dispersion $E^{1}_{\bm{k}}$. 
As mentioned previously, 
the field operator of the $m=1$ energy band can be given by 
the linear combination of $c^{a}_{\bm{k}\sigma}$ ($a=1,2,3$), 
as $a^{1}_{\bm{k}\sigma} = ( c^{1}_{\bm{k}\sigma} -\sqrt{2}c^{2}_{\bm{k}\sigma} +c^{3}_{\bm{k}\sigma} )/2$. 
Since $a=1,3$ and $a=2$ represent the OPs and IP, respectively, 
the $m=1$ energy band has larger component of the IP than OPs, 
so that more hole carriers are injected into the IP than OP. 
Thus, if the inter-layer single electron hopping arises as a form of Eq. (\ref{eq:H0}), 
more hole carriers tend to be introduced into the IP than OP. 
Furthermore, 
if the difference of the site potential between the OP and IP, which is now neglected in this tight-binding model, 
is taken into account, 
this tendency proceeds: 
this is because the site potential is higher in the IP than OP: $\Delta \varepsilon = \varepsilon_{\text{IP}} -\varepsilon_{\text{OP}} \sim 0.1$ eV. 

However, this is an opposite result to the experimental fact. 
Therefore the three-layer tight-binding model derived from the DFT calculations cannot describe 
the correct doping dependence of the OP and IP in the multi-layer cuprates.

\subsection{TPSC results} 
To obtain the correct doping dependence of the OP and IP, 
we consider the electron correlations in the multi-layer cuprates with the multi-layer TPSC approach. 

We first show the spin susceptibility in the OP and IP ($0 \leq q_{x,y} \leq \pi$), 
which are defined as $\chi^{\text{OP}}_{\text{S}} \equiv \chi^{11}_{\text{S}} = \chi^{33}_{\text{S}}$ 
and $\chi^{\text{IP}}_{\text{S}} \equiv \chi^{22}_{\text{S}}$, respectively. 
\begin{figure}[t] 
\begin{center} 
\includegraphics[width=8.0cm,clip]{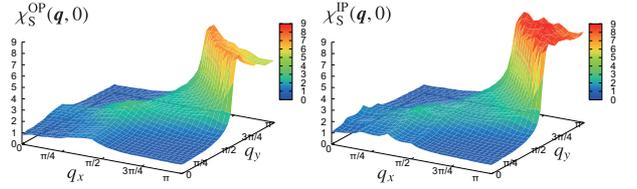} 
\caption{ 
Spin susceptibility in (Left) OP and (Right) IP ($0 \leq q_{x,y} \leq \pi$). 
The parameters are set to be $U=5.0$ eV, $n_{\text{av}}=0.85$, and $\epsilon_{m}=0$. 
} 
\label{fig:XS} 
\end{center} 
\end{figure} 
Here we have set the following parameters: 
the on-site Hubbard interaction $U=5.0$ eV, average filling $n_{\text{av}}=0.85$, 
and zero (Matsubara) frequency $\epsilon_{m}=0$. 
Similar to the non-interacting case, 
one can observe that the peak for $\chi^{\text{IP}}_{\text{S}}$ is larger than $\chi^{\text{OP}}_{\text{S}}$, 
so that the IP still have stronger AF instability than OP. 
This is also qualitatively consistent with the experimental fact suggested previously. 

\begin{figure}[t] 
\begin{center} 
\includegraphics[width=5.0cm,clip]{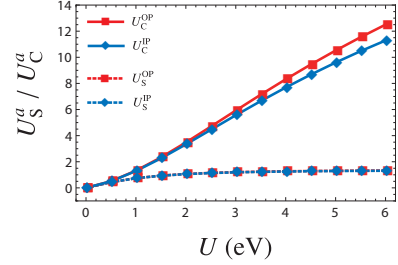} 
\caption{ 
Spin and charge channel interaction in the OP and IP at the average filing $n_{\text{av}}=0.85$. 
} 
\label{fig:USUC} 
\end{center} 
\end{figure}

\begin{figure}[t] 
\begin{center} 
\includegraphics[width=5.0cm,clip]{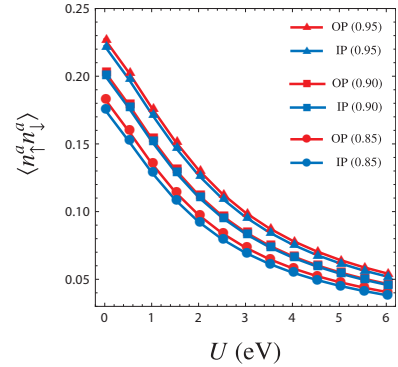} 
\caption{ 
Double occupancy in the OP and IP for $n_\text{{av}}= 0.85, \, 0.90, \, 0.95$. 
} 
\label{fig:nn} 
\end{center} 
\end{figure}

The spin and charge channel interaction, and double occupancy in the OP and IP can be determined self-consistently 
together with the spin (and charge) susceptibility, 
which are displayed in Fig. \ref{fig:USUC} and Fig. \ref{fig:nn}. 
% \begin{figure}[t] 
% \begin{center} 
% \includegraphics[width=8.0cm,clip]{UCUS_nn.eps} 
% \caption{ 
% (Left) Spin and charge channel interaction in the OP and IP at the average filing $n_{\text{av}}=0.85$. 
% (Right) Double occupancy in the OP and IP for $n_\text{{av}}= 0.85, 0.90, 0.95$. 
% } 
% \label{fig:USUC_nn} 
% \end{center} 
% \end{figure} 
The spin (charge) channel interaction $U^{a}_{\text{S}}$ ($U^{a}_{\text{C}}$) is suppressed (reinforced) in both of the OP and IP 
with increasing the on-site Hubbard interaction $U$. 
Then the contribution of the charge susceptibility $\chi_{\text{C}}$ to the self-energy $\Sigma$ gets decreased, 
so that the spin susceptibility $\chi_{\text{S}}$ mainly contributes to $\Sigma$. 
Also, the double occupancy is suppressed in both of the OP and IP, 
which is related to the spin channel in interaction $U^{a}_{\text{S}}$ through the TPSC ansatz Eq. (\ref{eq:ansatz}). 
Although the spin channel interaction is not so different between the OP and IP, 
the double occupancy in the IP gets small compared to the OP for each $n_\text{{av}}$. 
This means that larger energy gain in the IP can be obtained from the on-site Hubbard interaction $U$ than OP 
as a result from the strong electron correlations incorporated by the multi-layer TPSC approach. 

\begin{figure}[t] 
\begin{center} 
\includegraphics[width=6.0cm,clip]{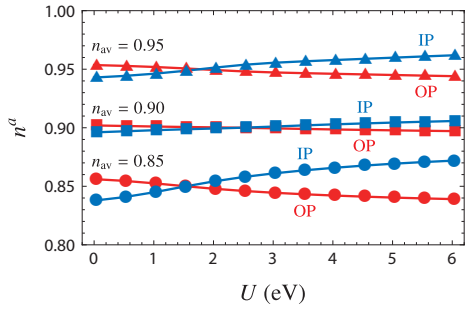} 
\caption{ 
Layer filling of the OP and IP $n^{a}$ for the average filling $n_{\text{av}}=0.85, \, 0.90, \, 0.95$ 
with varying the on-site Hubbard interaction $U$. 
} 
\label{fig:lf} 
\end{center} 
\end{figure}

Finally the layer filling $n^{a}$ is evaluated from the interacting Green's function 
through the self-energy derived from the multi-layer TPSC approach in Eq. (\ref{eq:self-energy}). 
In Fig. \ref{fig:lf} we show $n^{a}$ of the OP and IP for the average filling $n_{\text{av}}=0.85, \, 0.90, \, 0.95$ 
with varying the on-site Hubbard interaction $U$. 
Although for small $U$ situations the electron concentration in the OP large compared to IP, 
which is the same result as the non-interacting case as shown previously, 
more electrons tend to be introduced into the IP than OP with increasing $U$ for each $n_{\text{av}}$. 
This is because the double occupancy in the IP is small compared to OP, 
so that more energy gain can be obtained in the IP than OP from the on-site Hubbard interaction $U$ 
by moving electrons from the OP to IP. 
This result is consistent with the experimental fact, 
and this electron distribution between the OP and IP can be interpreted as a self-doping effect arising from strong electron correlations.

\section{Summary and Discussion} 
To summarize, 
considering a triple-layer Hubbard model as an effective model for the Hg-based triple-layer cuprate Hg-1223 
derived from the first principles calculations, 
we investigate it numerically with the multi-layer TPSC approach. 
We here show that the double occupancy (AF instability) decreases (increases) in the IP more than OP, 
and that the electrons tend to be introduced into the IP to obtain the energy gain from the on-site Hubbard interaction. 
These results are consistent with the experimental facts, 
and this electron distribution between the OP and IP can be interpreted as a self-doping effect arising from strong electron correlations. 

Although we have focused on an effective model of Hg-1223 as an example for the multi-layer cuprates, 
the self-doping effect may be considered to appear in the other triple-layer cuprates 
if their electron correlations are strong enough and their tight-binding parameters not so far from those for Hg-1223. 
Furthermore, 
this effect may also occur in the other multi-layer cuprates having $n > 3$ CuO$_2$ planes in a unit cell 
because it is attributed to 
the existence of the asymmetric CuO$_2$ planes 
and difference of the electron correlations in their CuO$_2$ planes. 

To obtain the energy gain from the on-site Hubbard interaction, 
the electron concentration in the IP tends to be approaching to the half-filling due to the self-doping effect. 
Of course, it does not reach the half-filling even when the different double occupancy occurs between the OP and IP. 
This is because, if too large difference of the layer filling between the OP and IP arises from the strong electron correlations, 
the energy gain cannot be obtained from the electron itinerancy, i.e., the inter-layer single electron hopping. 
Therefore the different layer-filling between the OP and IP gets balanced 
by the competition between the electron correlations and itinerancy 
before the layer-layer filling in the IP becomes the half-filling.  

% Since the electron concentration in the IP is approaching to haf-filling due to the self-doping effect, 
% the electron correlations become

\section*{Acknowledgment}
We would like to express my deepest gratitude to {\it AdvanceSoft Corporation}. 
I would also thank I. Maruyama, H. Sakakibara, T. Shirakawa, and S. Yunoki for their important advices and discussions. 
I wish to thank Kusakabe-lab members for their supports. 
The calculations were done in the computer centers of Kyushu University and ISSP, University of Tokyo. 
The work is supported by joint-project forgStudy of a simulation program for the correlated electron systemsh
with Advancesoft co. J161101009, and JSPS KAKENHI Grant Numbers JP26400357.

%\begin{acknowledgment}
%\acknowledgment
%I would like to express my deepest gratitude to {\it AdvanceSoft Corporation}. 
%I would also thank I. Maruyama, H. Sakakibara, T. Shirakawa, and S. Yunoki for their important advices and discussions. 
%I wish to thank Kusakabe-lab members for their supports. 
%The numerical calculations were done in the computer center of Kyushu University. 
%We wish to thank Shinichi Hikino, Toshihiro Sato, Kazuhiro Seki, and Hirofumi Sakakibara for useful discussions.   
% This study has been supported by Grants-in-Aid for Scientific Research from
% JSPS (Grants No. 23340095, R.A.). 
% R.A. acknowledges financial support from JST-PRESTO. 
%\end{acknowledgment}

% \appendix
% \section{}

\end{document}